\documentclass[twocolumn,twoside,slac_two]{revtex4}
\usepackage{amsfonts,amsmath,amssymb,bm,hyperref}
\usepackage{fancyhdr}

\begin{document}

\title{ON QUANTIZATION OF TIME-DEPENDENT SYSTEMS WITH CONSTRAINTS}

\author{ S.A. Gadjiev and R.G. Jafarov}
\affiliation{Institute for Physical Problems, Baku State
University, AZ 1148, Baku, Azerbaijan} \email{jafarov@hotmail.com}

\newcommand{\ba}{\begin{eqnarray}}
\newcommand{\ea}{\end{eqnarray}}
\newcommand{\tr}{\,\mbox{tr}\,}

\begin{abstract}
The Dirac method of canonical quantization of theories with second
class constraints has to be modified if the constraints depend on
time explicitly. A solution of the problem was given by Gitman and
Tyutin. In the present work we propose an independent way to
derive the rules of quantization for these systems, starting from
physical equivalent theory with trivial non-stationarity.
\end{abstract}

\maketitle

\section{\bf Introduction}

In relativistic particle theories and string theories explicit
time-dependent gauges are often used \cite{gt1}. Not only this
problem but also others are bringing out the necessity to
formulate in general rules of quantization of time-dependent
systems with constraints. The canonical quantization of
time-dependent systems with constraints has been formulated by
Dirac \cite{d} and described in \cite{gt2}.The generalization of
the Dirac method of canonical quantization for the case of
time-dependent constraints was described in the book of Gitman and
Tyutin \cite{gt2}. A development of the method and examples can be
found in \cite{gg}. In this paper the interpretation of two
general moments on which Gitman-Tyutin quantization
(GT-quantization) is based is given. These are: formal
introduction of a momentum $\epsilon $ conjugated to the time $t$,
and postulation of a special non-unitary
Schr$\stackrel{..}{o}$dinger time dependent of operators.

\section{\bf GT- quantization of theories with time-dependent second-class constraints}

Here, we briefly describe the modification of the Dirac method of
quantization for time-dependent second-class constraints proposed
in \cite{gt2}.

\indent Let us have a theory in a Hamiltonian formulation with
second-class constraints $\Phi (\eta,t)=0$, $\eta =(q,p)$, which
can explicitly depend on the time $t$. Then the equation of motion
of such a system may be written in the usual form, if one formally
introduces a momentum $\epsilon $ conjugated to the time $t$, and
defines the Poisson bracket in the extended space of canonical
variables $(q, p, t, \epsilon)=(\eta, t, \epsilon)$,

\begin{equation}
\stackrel{.}{\eta }=\left. \left\{ \eta ,\,H+\epsilon \right\}
\right| _{D(\Phi )},\,\,\, \Phi (\eta ,t)=0 , \label{1}
\end{equation}
where $H$ is a Hamiltonian of the system, and $\{{A,
B}\}_{D(\Phi)}$ is the notation for the Dirac bracket with respect
to a system of second-class constraints $\Phi$. The Poisson
bracket, wherever encountered, is henceforth understood to be one
in such an above mentioned extended space. The total derivative of
an arbitrary function $A(\eta, t)$, with allowance made for the
equations (1), has the form

$$
\frac{dA}{dt}=\left. \left\{ A,\,H+\epsilon \right\} \right| _{D(\Phi )}.
$$

\indent In this case the quantization procedure in the
Schr$\stackrel{..}{o}$dinger picture can be formulated as follows.
The variables $\eta$ of the theory are assigned the operators
$\stackrel{\wedge }{\eta}$, which satisfy the equal-time
commutation relations ( $ [,\}$ denotes the generalized
commutator, commutator or anti-commutator depending on the
parities of the variables),
\begin{equation}
[ \stackrel{\wedge }{\eta },\stackrel{\wedge }{\eta ^\prime}  \}
=\left. i\{ \eta ,\eta ^{\prime } \} _{D(\Phi )}\right| _{\eta =
{\stackrel{\wedge}{\eta }}},
\end{equation}
the constraints equation
$$
\Phi (\stackrel{\wedge }{\eta },t)=0,
$$
and equations of evaluation ( we disregard problems connected with
operator ordering)
\begin{equation}
\stackrel{\stackrel{\cdot }{\wedge }}{\eta }=-\left. \left\{ \eta
,\epsilon \right\} _{D(\Phi )}\right| _{\eta =\stackrel{\wedge }{\eta }%
}=\left. -\left\{ \eta ,\Phi _l\right\} \left\{ \Phi ,\Phi \right\}
_{l,l^{^{\prime }}}^{-1}\frac{\partial \Phi _{l^{^{\prime }}}}{\partial t}%
\right| _{\eta =\stackrel{\wedge  }{\eta }}.
\end{equation}

\indent To each physical quantity $A$ given in the Hamiltonian
formalism by the function $A(\eta , t)$, we assign a
Schr$\stackrel{..}{o}$dinger operator $\stackrel{\wedge }{A}$ by
the rule $\stackrel{\wedge }{A}=A(\stackrel{\wedge }{\eta },t)$;
in the same manner we construct the quantum Hamiltonian
$\stackrel{\wedge }{H}$, according to the classical Hamiltonian
$H(\eta , t)$. The time evaluation of the state vector $\psi$ in
the Schr$\stackrel{..}{o}$dinger picture is determined by the
Schr$\stackrel{..}{o}$dinger  equation

\begin{equation}
i\frac{\partial \psi }{\partial t}=\stackrel{\wedge }{H}\psi ,\,\,\stackrel{
\wedge }{H}=H(\stackrel{\wedge }{\eta },t).
\end{equation}
From (3) it follows, in particular, that
$$
\frac{d{\stackrel{\wedge }{A}}}{dt}=\left. \left\{ A,\epsilon
\right\} _{D(\Phi )}\right| _{\eta =\stackrel{\wedge }{\eta }}
$$
and, as a consequence of (2), for arbitrary
Schr$\stackrel{..}{o}$dinger operators $\stackrel{\wedge }{A }$,
$\stackrel{\wedge }{B }$ we have

$$
[\stackrel{\wedge }{A},\stackrel{\wedge }{B}\}=\left.
i\{A,B\}_{D(\Phi )}\right| _{\eta =\stackrel{\wedge }{\eta }}.
$$

\indent It is possible to see that quantum theories, which
correspond to different initial data for the equation (3), are
equivalent.

\indent One can adduce some arguments in favor of the proposed
quantization procedure. For instance, to check that the
correspondence principle between classical and quantum equations
of motion holds true in this procedure, we pass over to the
Heisenberg representation, whose operators $\stackrel{\vee  }{\eta
}$ are related to the operators $\stackrel{\wedge  }{\eta }$ as
$\stackrel{\vee }{\eta }=U^{-1}\stackrel{\wedge }{\eta }U$, where
$U$ is the operator of the evolution of the
Schr$\stackrel{..}{o}$dinger equation,
$$
i\frac{\partial U}{\partial t}=\stackrel{\wedge }{H}U,\,\,\,\left.
U\right| _{t=0}=1.
$$
Heisenberg operator $\stackrel{\vee }{A}$ of an arbitrary physical
quantity $A$ is constructed from the corresponding
Schr$\stackrel{..}{o}$dinger operator $\stackrel{\wedge }{A}$ in
the same manner $\stackrel{\vee }{A}=U^{-1}\stackrel{\wedge
}{A}U$. One can find the total time derivative of the Heisenberg
operator $\stackrel{\vee }{A}$:

\begin{equation}
\frac{d\stackrel{\vee }{A}}{dt}=\left. \{A,H+\epsilon \}_{D(\Phi )}\right|
_{\eta =\stackrel{\vee }{\eta }},
\end{equation}
which coincides in form with the classical equation of motion.

\indent It follows from (5) that the
Heisenberg operators $\stackrel{\vee }{\eta }$ also satisfy the equation

\begin{equation}
\stackrel{\stackrel{\cdot }{\vee }}{\eta }=\left. \{\eta ,H+\epsilon
\}_{D(\Phi )}\right| _{\eta =\stackrel{\vee }{\eta }}.
\end{equation}
Besides, one can easily verify that the equal-time relations hold
for these operators,
$$
[\stackrel{\vee }{\eta },\stackrel{\vee }{\eta }^{\prime
}\}=\left.
i\{\eta ,\eta ^{\prime }\}_{D(\Phi )}\right| _{\eta ={\stackrel{\vee }{\eta }}%
},\,\,\, \Phi (\stackrel{\vee }{\eta },t)=0.
$$
These relations, together with (6), may be regarded as a
prescription of the quantization in the Heisenberg picture for
theories with time-dependent second-class constraints.

\indent Note that the time dependence of Heisenberg operators in
the theories considered is not unitary in the general case. In
other words, no such (Hamiltonian) operator exists, whose
commutator with a physical quantity would give its total time
derivative. This is explained by the existence of two factors
which determine the time evolution of the Heisenberg operator. The
first one is the unitary evolution of the state vector in the
Schr$\stackrel{..}{o}$dinger picture, while the second one is the
time variation of Schr$\stackrel{..}{o}$dinger operators
$\stackrel{\wedge }{\eta }$, which in the general case is
non-unitary. Physically, this is explained by the fact that the
dynamics develops on a surface which itself changes with time.

\section{\bf Alternative approach to quantization of systems with time-dependent
constraints}
Let us show that the equation (1) arises naturally
from a consideration of a modified formulation, which is trivial
non-stationary.

\indent  Let $L=L(q,\stackrel{\cdot }{q},t)$ be a time-dependent
Lagrangian of some singular theory,
($q=(q_1,...,q_n),\,\stackrel{\cdot }{q=\frac{dq}{dt}}$). One can
consider another Lagrangian $L^{\prime }=L^{\prime
}(q,\stackrel{\cdot }{\stackrel{\cdot }{q},\tau ,\zeta ,t})$,
which depends on two supplementary variables $\tau, \zeta $ and
connected with origin Lagrangian $L$ as:

\begin{equation}
L^{\prime }=\stackrel{\sim }{L}+\zeta (\tau -t),\,\,\,
\stackrel{\sim } {L}=L(q,\stackrel{\cdot }{q},\tau ).
\end{equation}
The theory with Lagrangian $L^{\prime }$ is equivalent to the
theory with Lagrangian $L$ in the sector of variables $q$. Indeed,
the Lagrange equations in the theory with $L^{\prime }$ have the
form:

\begin{equation}
\frac{\delta L^{\prime }}{\delta q}=\frac{\partial \stackrel{\sim }{L}}{%
\partial q}-\frac d{dt}\frac{\partial \stackrel{\sim }{L}}{\partial
\stackrel{\cdot }{q}}=0,
\end{equation}

\begin{equation}
\frac{\delta L^{\prime }}{\delta \tau}=\zeta +\frac{\partial \stackrel{\sim }{%
L}}{\partial \tau }=0,\,\,\,\,\frac{\delta L^{\prime }}{\delta
\zeta }=\tau -t=0.
\end{equation}
Taking (9) into account in (8), it is easy to derive the equations:

$${\frac{\delta L}{\delta q}}=\frac{\partial L}{\partial q}-\frac d{dt}\frac{%
\partial L}{\partial \stackrel{\cdot }{q}}=0.
$$

\indent Let us consider the Hamiltonian formulation of the theory with
Lagrangian $L^{\prime }$. Introduce momenta

\begin{equation}
p=\frac{\partial L^{\prime }} {\partial
\stackrel{.}{q}}=\frac{\partial \stackrel{\sim }{L} }{\partial
\stackrel{.}{q}},\,\,\,\,\,\,\,\epsilon =\frac{\partial L^{\prime
}}
{\partial \stackrel{.}{%
\tau }}=0,\,\,\,k=\frac{\partial L^{\prime }}{\partial
\stackrel{.}{\zeta }}=0 .
\end{equation}
From the relation (10) one can find primarily expressible
velocities $\stackrel{.}{X}$ and primary constraints $\Phi^{(1)}$,
($q=(X,x),\,\,\stackrel{\cdot }{x}=\lambda $),\,\,$\stackrel{\cdot
}{X}=V(q,p,\lambda ,\tau )$,\,\, $\stackrel{\sim }{\Phi
}^{(1)}=0$, where  $\stackrel{\sim }{\Phi }^{(1)}=\Phi
^{(1)}(q,p,\tau )$. $\Phi ^{(1)}(q,p,t )$ are constraints in the
theory with Lagrangian $L$. Then Hamiltonian $H^{(1)^\prime} $ has
the form:
$$
H^{(1)^{\prime }}=(\frac{\partial L^{\prime }}{\partial \stackrel{\cdot }{q}}%
\stackrel{\cdot }{q}-\stackrel{\sim }{L}-\zeta (\tau -t))_{X=V}=\stackrel{%
\sim }{H}-\zeta (\tau -t)
$$
\begin{equation}
+\lambda \stackrel{\sim }{\Phi }^{(1)}+\lambda _\epsilon \epsilon
+\lambda _kk,
\end{equation}
where $\stackrel{\sim }{H}=H(q,p,\tau )$. $H(q,p,t)$ is the
Hamiltonian of a theory with Lagrangian $L$.

\indent The condition of conservation of constraint $k=0$ in time
gives

$$
\stackrel{\cdot }{k}=\{k,H^{(1)\prime }\}=-\frac{\partial H^{(1)\prime }}{%
\partial \zeta }=\tau -t=0.
$$
Thus, the secondary constraint ${\Phi }_1^{(2)}=\tau -t$ appears.
Considering a condition of its conservation, we define $\lambda
_\epsilon $:

$$
\stackrel{\cdot }{\Phi }_1^{(2)}=\frac{\partial \Phi
_1^{(2)}}{\partial t}+\{\Phi _1^{(2)},H^{(1)\prime }\}=-1+\lambda
_\epsilon =0,\,\,\,\,\lambda _\epsilon =1.
$$
From the condition of the constraint $\epsilon=0$ we get

$$
\stackrel{\cdot }{\epsilon }=\{\epsilon ,H^{(1)\prime
}\}=-\frac{\partial H^{(1)\prime }}{\partial \tau
}=-\frac{\partial \stackrel{\sim }{H}}{\partial \tau}+\zeta
-\lambda \frac{\partial \stackrel{\sim }{\Phi }^{(1)}}{\partial
\tau }=0,
$$
or $\zeta =f(q,p,\tau ,\lambda )$. From the condition of this
constraint conservation in time we find $\lambda _k=\varphi
(q,p,\lambda ,\stackrel{\cdot }{\lambda })$, where $f$ and
$\varphi $ are some functions.

\indent Substituting the Lagrange multipliers found in the
Hamiltonian (11), one transforms it to the form:

\begin{equation}
H^{(1)\prime }=\stackrel{\sim }{H}-\zeta (\tau -t)+\lambda \stackrel{\sim }{%
\Phi }^{(1)}+\varphi k+\epsilon.
\end{equation}

\indent Further, one can continue the Dirac procedure to obtain
$\lambda $- multipliers and secondary constraints already with
Hamiltonian (12) [3]. Using the Dirac procedure for the
constraints $\stackrel{\sim }{\Phi }^{(1)}$, one can use the
following Hamiltonian
$$
H^{(1)\prime }=\stackrel{\sim }{H}_{eff}+\lambda \stackrel{\sim }{\Phi }%
^{(1)},
$$
instead of the Hamiltonian (12), where $\stackrel{\sim
}{H}_{eff}=\stackrel{\sim }{H}+\epsilon $. This is so because of
the constraints $\stackrel{\sim }{\Phi }^{(1)}$ and secondary
constraints, which can appear, do not contain the variables $\zeta
$ and $k$. Let $\stackrel{\sim }{\Phi }^{(2)}$ be secondary
constraints. Then,
$$
\stackrel{\sim }{\Phi }^{(2)}=\Phi ^{(2)}(q,p,\tau ),
$$
where $\Phi ^{(2)}(q,p,t)$ are secondary constraints of the theory
with Lagrangian $L$.

\indent Suppose that $\Phi =(\Phi ^{(1)},\Phi ^{(2)})$ is a total
system of constraints of the theory with Lagrangian $L$ and is of
second class. Then the total system of constraints of the theory
with Lagrangian $L^{\prime }$ is also of second class and has the
form
$$
\stackrel{\sim }{\Phi }=0,\,\,\,\,\tau -t=0,\,\,\,\,\epsilon =0,
$$
$$
\zeta -f=0,\,\,\,\,k=0,\,\,\,\,\stackrel{\sim }{\Phi }=(\stackrel{\sim }{%
\Phi }^{(1)},\stackrel{\sim }{\Phi }^{(2)}).
$$
Because the constraints for the variables $\zeta $ and $k$ have
the special forms [3], they can be merely excluded from the
equations of motion by means of the constraints. In doing this,
one can easily discover that the equations for the rest variables
coincide with equation (1).

\indent The quantization in Schr$\stackrel{..}{o}$dinger
picture of the theory in question may be formulated
in the following form. We have a Hamiltonian theory with
canonical variables $\eta =(q,p)$ and $(t,\epsilon )$.
The surface of constraints is described by the equation
$$
\Phi (\eta ,t)=0.
$$
The Hamiltonian of the theory is $H$. When quantizing, all
variables became operators with commutation relations

\begin{equation}
[\stackrel{\wedge }{Q},\stackrel{\wedge }{Q}^\prime \}=\left.
i\{Q,Q^{\prime }\}_{D(\Phi )}\right| _{Q=\stackrel{\wedge }{Q}%
}\,,\,\,\,\,\,Q=(\eta ;t,\epsilon ).
\end{equation}
Constraints are equal to zero,
$$
\Phi (\stackrel{\wedge }{\eta },t)=0,
$$
and conditions on the state-vectors hold (similar to the Dirac
quantization of theories with first class constraints [2]),
\begin{equation}
H_{eff}\psi =0\,\,,\,\,\,\,H_{eff}=H+\epsilon .
\end{equation}

\indent One can verify that this quantization is fully equivalent
to the GT-quantization in the Schr$\stackrel{..}{o}$dinger picture.
Indeed, let us realize $\stackrel{\wedge }{t}$ as operator of
multiplication by the variable $t$. Then,
$$
\stackrel{\wedge }{\epsilon }=-i\frac \partial {\partial t}.
$$
From (11) we have
\begin{equation}
[\stackrel{\wedge }{\eta },\stackrel{\wedge }{t}]_{-}=0\,;\,\,\,\,[\stackrel{%
\wedge }{\eta },\stackrel{\wedge }{\epsilon }]_{-}=\left. i\{\eta
,\epsilon \}_{D(\Phi )}\right| _{\eta =\stackrel{\wedge }{\eta }}.
\end{equation}
And in the selected realization,
\begin{equation}
[\stackrel{\wedge }{\eta },\stackrel{\wedge }{\epsilon }]_{-}=i\stackrel{%
\wedge }{\tau },
\end{equation}
the condition (3) follows from (15) and (16). Finally, the
condition (14) is the Schr$\stackrel{..}{o}$dinger equation.

\end{document}